\documentclass[11pt]{revtex4}
\usepackage{amssymb,epsf}
\usepackage{latexsym}
\begin{document}

\title{MOND Theory and Thermodynamics of Spacetime}

\author{Ahamd Sheykhi$^{1,2}$ and Leila Liravi$^{1}$
\footnote{email:asheykhi@shirazu.ac.ir}}
\affiliation{$^{1}$Department of Physics, College of Science, Shiraz University, Shiraz 71454, Iran\\
         $^{2}$Biruni Observatory, College of Science, Shiraz University, Shiraz 71454, Iran}

\begin{abstract}
Starting from the Modified Newtonian Dynamics (MOND) theory and
using an inverse approach, we construct a general form of the
entropy expression associated with the horizon based on the
entropic nature of gravity. Using the thermodynamics-gravity
correspondence in the cosmological setup, we apply the corrected
entropy expression and find the modified Friedmann equation by
three methods, namely, (i) the first law of thermodynamics, (ii)
the entropic force scenario and (iii) the emergence nature of
gravity. We confirm that our model guaranties the generalized
second law of thermodynamics for the universe enveloped by the
apparent horizon. Our studies reveal that the MOND theory of
gravity may be naturally deduced from the modification of the
horizon entropy. These results may fill in the gap in the
literatures, understanding the theoretical origin of the MOND
theory from thermodynamics-gravity conjecture.

\end{abstract}

\maketitle

\section{Introduction\label{Intro}}
Two main challenges of the modern cosmology are the so called
\textit{dark matter} puzzle and \textit{dark energy} problem. The
former originates from the fact that the total luminous mass of
galaxies and clusters of galaxies at large scales, does not
provide sufficient gravitation to explain the observed dynamics of
these systems. To overcome the problem, that is, to explain the
rotation curves of spiral galaxies or the dynamics of the clusters
of galaxies, one needs to consider an extra component of mass,
which is uniformly distrusted around the galaxies and provides the
necessary gravitation. However, one may argue that the problem is
due to the flaw in the Newton's law of gravitation at large
scales. In this direction, alternative theories of gravitation
have been speculated and debated. Among them, the so called MOND
theory \cite{Milgrom1} is widely accepted, although its
theoretical origin is still doubtful. Many attempts have been done
to address the theoretical origin of the MOND theory. For example,
in \cite{Limond,Sheymond}, the authors argued that the origin of
the MOND theory can be understood from Debye entropic gravity
scenario. Other studies to disclose the origin of the dark matter
puzzle have been carried out in \cite{Sunny1,Sunny2,Sunny3}. In
particular, very recently, it was suggested that primordial
regular black holes produced during inflation can be regarded as
the source of the dark matter \cite{Sunny4,Sunny5}.

The latter comes from the fact that the observations of type Ia
supernova explosions in high redshift galaxies confirm that our
universe is currently undergoing a phase of accelerated expansion.
This was an unexpected discovery and shook the foundations of the
modern cosmology.  In the context of standard cosmology, one needs
to consider an extra unknown component of energy, which is usually
called in the literatures as dark energy. What we know from dark
energy is that it is smoothly filled all spaces and has
anti-gravity nature which push our universe to accelerate.
However, there is another way to justify the acceleration of the
cosmic expansion, namely taking into account alternative theories
of gravity such as $f(R)$ gravity, Gauss-Bonnet gravity, or brane
cosmology, etc.

On the other side, in recent years, there has been more
theoretical progress on understanding the nature of gravity. It
was argued that when the spacetime, as  a large scales system, is
considered as a thermodynamical system, then there is a profound
connection between the laws of gravity describing the spacetime
geometry and the laws of thermodynamics. The correspondence
between the gravitational field equations and thermodynamics has
been disclosed in three levels. At the first level, it was shown
by Jacobson \cite{Jac} that the hyperbolic second order partial
differential Einstein equation for the spacetime metric has a
predisposition to thermodynamic behavior. He disclosed that the
gravitational Einstein equation can be derived from the relation
between the horizon area and entropy, together with the Clausius
relation $\delta Q=T\delta S$ \cite{Jac}. The correspondence
between the first law of thermodynamics and gravitational field
equations has been extended to $f(R)$ gravity \cite{Elin},
Gauss-Bonnet gravity, the scalar-tensor gravity and more general
Lovelock gravity \cite{Pad1,Pad2,Cai1,Pad3}. In the context of
Friedmann-Robertson-Walker (FRW) cosmology it has been confirmed
that the first Friedmann equation on the apparent horizon can be
translated to the first law of thermodynamics, $dE=TdS+WdV$, and
vice versa \cite{wang1,wang2,CaiKim,Cai2,Shey1,Shey2,Shey3,Shey4}.
The correspondence between the first law of thermodynamics on the
boundary and the gravitational field equations in the bulk sheds
also the light on holography. These results further support the
idea that gravitation on a macroscopic scale is a manifestation of
thermodynamics. At the deeper level, it was shown that gravity is
not a fundamental force and can be treated as an emergence
phenomenon. Starting from the first principles, namely the
holographic principle and the equipartition law of energy on the
horizon degrees of freedom, Verlinde \cite{Ver} argued that the
change in the information of the system leads to an entropic force
which can be translated to the law of gravity. The entropic nature
of gravity has been widely explored (see e.g.
\cite{Cai3,sheyECFE,Visser} and references therein). In both
approaches mentioned above one considers the spacetime as a
pre-exist geometry. Is it possible to regard the spacetime itself
as an emergent structure? In the deepest level, Padmanabhan
proposed that the spatial expansion of the universe can be
understood as the consequence of the emergence of space
\cite{PadEm}. According to Padmanabhan's proposal  the
\textit{cosmic space emerges as the cosmic time progress}. In this
approach the most fundamental notion namely the degrees of freedom
of the matter fields in the bulk and the degrees of freedom on the
boundary play crucial role. Indeed, by counting the difference
between degrees of freedom on the boundary and in the bulk and
equating it with the volume change, one is able to construct the
dynamical equations describing the evolution of the universe
\cite{CaiEm,Yang,Sheyem}.

Independent of the approach for dealing with the
thermodynamics-gravity conjecture, the entropy expression
associated with the boundary of system, plays a crucial role in
extracting the gravitational field equations from thermodynamic
arguments. In the cosmological background, any modification to the
entropy associated with the apparent horizon of FRW universe,
implies a modification to the Friedmann equations which leads to a
modified cosmology
\cite{CaiLM,SheT1,Odin1,SheT2,Emm2,SheB1,SheB2,Odin2,Odin3,Odin4,Odin5}.
Some of these modified cosmological models, inspired by
thermodynamics-gravity correspondence, can explain the late-time
acceleration of the cosmic expansion without invoking additional
component of energy \cite{SheT2,Emm2}.

In the present work, we are going to construct a general form of
the entropy associated with the boundary, which may simultaneously
address both flat rotation curves of spiral galaxies and the late
time accelerated expansion. For this purpose, we start from a
general expression for the MOND theory. Using an inverse approach,
as well as the entropic force scenario for the MOND theory, we
reconstruct a general form of the entropy. Applying the obtained
entropy expression to the cosmological setup, we are able to
extract the modified Friedmann equations by using
thermodynamics-gravity conjecture. We shall also check the
validity of the generalized second law of thermodynamics for the
universe enveloped by the apparent horizon.

This paper is structured as follows. In section \ref{MOND}, we use
the entropic nature of gravity and start from the MOND theory to
construct the general form of the entropy associated with the
horizon. In section \ref{FIRST}, we start from the first law of
thermodynamics on the apparent horizon and apply the modified
entropy expression to establish corrections to the Friedmann
equations. Given the general form of the entropy inspired by MOND
theory, in section \ref{Entropic}, we apply the entropic force
scenario to construct the modified Friedmann equations. In section
\ref{Emergence}, we use the idea of emergence gravity and
reconstruct the modified Friedmann equations. In section
\ref{GSL}, we confirm that our cosmological model guaranties the
generalized second law of thermodynamics for the universe
enveloped by the apparent horizon. The last section is devoted to
the closing remarks.
\section{Corrections to entropy inspired by MOND theory\label{MOND}}
The Modified Newtonian dynamics (MOND) suggested by Milgrom to
explain the flat rotation curves of the spiral galaxies
\cite{Milgrom1}. According to the MOND theory, the Newton's second
law get modified for the large scales as
\begin{equation}\label{F0}
F=m  \mu \left(\frac{a}{a_0}\right)a,
\end{equation}
where $a$ stands for the usual kinematical acceleration, which is
taken as $a=v^{2}/R$, and $a_{0}=1.2\pm 0.27  \times 10^{-10}$
$m/s^{2}$ is a constant \cite{Begeman}. Here $\mu(x)$ is a real
function satisfies the following boundary conditions,
\begin{equation}\label{mu}
\mu(x)\approx\left\{
  \begin{array}{ll}
  $$1$$  \quad \quad {\rm for} \quad\quad {x}\gg 1, &  \\
  $$x$$  \quad \quad {\rm for}  \quad\quad {x}\ll 1.
    \end{array}
\right.
\end{equation}
At large distance, at the galaxy out skirt, the kinematical
acceleration `$a$' is extremely small, smaller than $10^{-10}$
$m/s^{2}$ , i.e., $a\ll a_{0}$, hence the function $\mu
(\frac{a}{a_{0}})=\frac{a}{a_{0}}$. Mathematically, for a galaxy
with mass $M$ and a star (particle) with mass $m$, the Newton's
law of gravity get modified as
\begin{equation}\label{v}
F=m \frac{a^2}{a_{0}}=\frac{GMm}{R^2}, \Rightarrow  v= (GM
a_{0})^{1/4}\approx cte.
\end{equation}
This implies that the velocity of star, on circular orbit from the
galaxy-center is constant and does not depend on the distance; the
rotational-curve is flat, as it observed.

Our aim here is to ansatz a general form for  $\mu(x)$ which
satisfies conditions (\ref{mu}). Among several function which may
satisfy condition (\ref{mu}), and motivated with the previous
studies in this direction, we propose the following function,
\begin{eqnarray}\label{mu1}
\mu(x)=x(1+x^{\alpha})^{-1/{\alpha}},
\end{eqnarray}
where $\alpha>0$. Note that the interpolation function has usually
been given the following functional form \cite{Begeman,Gentile}
\begin{eqnarray}\label{mualph2}
\mu(x)=\frac{x}{\sqrt{1+x^2}}.
\end{eqnarray}
However, an alternative simple interpolating function,
\begin{eqnarray}\label{mualph1}
\mu(x)=\frac{x}{1+x},
\end{eqnarray}
was also propped, which provides a less sudden transition from the
Newtonian to the MOND regime than does the standard function
\cite{Famaey1,Famaey2}. It is clear that the general form we
proposed in (\ref{mu1}) reduces to functions (\ref{mualph2}) and
(\ref{mualph1}) in the limiting case where $\alpha=1,2$. One can
easily check that (\ref{mu1}) satisfies both conditions
(\ref{mu}). On the other hand, since $\alpha>0$ is a free
parameter, one can expand expression (\ref{mu1}), and reproduce
all terms in the series expansion. If one choose other types of
function $\mu(x)$, their expansions may be similar to the proposed
function (\ref{mu1}), by suitably choosing the free parameter
$\alpha$. This discussion may justify our  ansatz  for function
(\ref{mu}).

We consider a system that its boundary is not infinitely extended
and forms a closed surface with spherical geometry. We can take
the boundary as a storage device for information, i.e. a
holographic screen. We also assume at the center of the
holographic screen there is a mass $M$ and at distance $R$, mass
$m$ is located near the screen. Using the entropic force scenario
\cite{Ver}, and taking into account a general form for the entropy
associated with the holographic screen, we can write down the
Newton's law of gravitation as  ($k_B=\hbar=c=1$) \cite{sheyECFE}
\begin{eqnarray}\label{F2}
F&=&\frac{GMm}{R^2}\times 4G  \frac{ dS_h}{dA}\mid _{A=4\pi
R^2}\nonumber\\&&=4 G m  a \times \frac{ dS_h}{dA}\mid _{A=4\pi
R^2},
\end{eqnarray}
where $a=GM/R^2$ is the acceleration of a particle with mass $m$
which rotates at the distance $R$ around the central mass $M$.
Equating expression  (\ref{F2})  with Eqs. (\ref{F0}), after using
(\ref{mu1}), we find
\begin{equation}\label{mu2}
\mu(x)=4G \frac{ dS_h}{dA}.
\end{equation}
The key point here is to recognize in Eq. (\ref{mu1}),
$x=a/a_{0}$, with $a_{0}=\gamma \pi M$, where $\gamma$ is a
parameter which can be constrained by observation. This implies
that we can write $x=4G/(\gamma A)$, where $A=4\pi R^2$. Clearly,
$x$ can be calculated for each galaxy or cluster of galaxies.
Since $M$ is, at least, of order of a galaxy mass and $a_{0}\simeq
10^{-10} m/s^2\ll 1$, hence $\gamma\ll 1$. In terms of the horizon
area, $A$, the functional form of $\mu$ can be written as
\begin{equation}\label{mu3}
\mu(A)=\Big{\{}1+\left(\frac{\gamma
A}{4G}\right)^{\alpha}\Big{\}}^{-1/{\alpha}}.
\end{equation}
Integrating Eq. (\ref{mu2}), we find the entropy associated with
the horizon as
\begin{equation}\label{Sh2}
S_h=\frac{1}{4G}\int{\mu(A)dA}=\frac{1}{4G}\int{\Big{\{}1+\left(\frac{\gamma
A}{4G}\right)^{\alpha}\Big{\}}^{-1/{\alpha}}dA}.
\end{equation}
The integral can be done in terms of hypergeometric function,
$_2F_1(a,b,c,z)$, and can be written in a compact form. The result
is
\begin{equation}\label{Sh3}
S_h=\frac{A}{4G} \times \text{{\ }} _{2}F_{1}\Big{\{}
\frac{1}{\alpha},\frac{1}{\alpha},\frac{\alpha+1}{\alpha}, -\left(
\frac{\gamma A}{4G}\right)^{\alpha}\Big{\}}.
\end{equation}
For $\alpha=1$, the above expression for entropy reduces to
\begin{equation}\label{ShRenyi}
S_h=\frac{1}{\gamma}\ln \left(1+\frac{\gamma A}{4G}\right),
\end{equation}
which is the well-known Renyi entropy \cite{Renyi}. On the other
hand, for $\alpha=2$, expression (\ref{Sh3}) restores
\begin{equation}\label{ShKan}
S_h=\frac{1}{\gamma} \ln \left(\frac{\gamma
A}{4G}+\sqrt{1+\left(\frac{\gamma A}{4G}\right)^2}\right),
\end{equation}
which is a deformed (dual) version of the Kaniadakis entropy
\cite{AbreuNet1}. Using the fact that $_2F_1(a,b,c,z)$ has a
convergent series expansion for $|z| <1$, we can expand expression
(\ref{Sh3}), up to linear term in $\eta$, as

\begin{equation}\label{Sh4}
S_h=\frac{A}{4G} \Big{\{}1-\eta \left(\frac{
A}{4G}\right)^{\alpha}+...\Big{\}},
\end{equation}
where
\begin{equation}
\eta=\frac{\gamma^{\alpha}}{\alpha(\alpha+1)}\ll 1.
\end{equation}
Finally, let us emphasize that although in Eq. (\ref{mu}),
parameter $\alpha>0$ can be any positive number, but for
simplicity, in the remaining part of this work we assume $\alpha$
is a positive integer number, although one can relax this
assumption.
\section{Modified Friedmann equations from the first law of thermodynamics\label{FIRST}}
We consider a spatially homogeneous and isotropic universe with
line elements
\begin{equation}
ds^2={h}_{\mu \nu}dx^{\mu} dx^{\nu}+R^2(d\theta^2+\sin^2\theta
d\phi^2),
\end{equation}
where $R=a(t)r$, $x^0=t, x^1=r$, the two dimensional metric is
given by $h_{\mu \nu}$=diag $(-1, a^2/(1-kr^2))$, and $k = -1,0,
1$, stands for open, flat, and closed universes, respectively. The
dynamical apparent horizon, a marginally trapped surface with
vanishing expansion, is determined by the relation $h^{\mu
\nu}\partial_{\mu}R\partial_{\nu}R=0$, which implies that the
vector $\nabla R$ is null on the apparent horizon surface. For FRW
geometry, the explicit evolution of the apparent horizon radius
reads \cite{Hay1,Hay2,Bak}
\begin{equation}
\label{radius}
 R=\frac{1}{\sqrt{H^2+k/a^2}},
\end{equation}
where $H=\dot{a}/a$ is the Hubble parameter. The apparent horizon
is a suitable boundary from thermodynamic arguments \cite{Cai2}.
The surface gravity in a dynamical background should, in
principle, include contributions from the time derivative of the
apparent horizon radius. In this context, the Hayward-Kodama
surface gravity, defined by \cite{Hay1,Hay2,Bak}
\begin{equation}\label{kappa}
\kappa = \frac{1}{2\sqrt{-h}} \partial_{\mu} \left( \sqrt{-h}
h^{\mu \nu}
\partial_{\nu} R \right),
\end{equation}
which is indeed the most general and invariant definition for
surface gravity associated with a dynamical apparent horizon. Note
that surface gravity includes terms involving both the Hubble
parameter and its time derivative. The temperature associated with
the dynamical apparent horizon is then defined as
\cite{Hay1,Hay2,Bak}
\begin{equation}\label{T}
T_h=\frac{\kappa}{2\pi}=-\frac{1}{2 \pi R}\left(1-\frac{\dot
{R}}{2HR}\right).
\end{equation}
To avoid negative temperature one can also define
$T=|\kappa|/2\pi$. Besides, when $\dot {R}\ll 2HR$, which
physically means that the radius of the apparent horizon is almost
fixed, one may define $T=1/(2\pi R )$ \cite{CaiKim}. We further
assume the energy content of the universe is in the form of
perfect fluid with energy-momentum tensor
$T_{\mu\nu}=(\rho+p)u_{\mu}u_{\nu}+pg_{\mu\nu},$ where $\rho$ and
$p$ are the energy density and pressure, respectively. As far as
we know, there is no energy exchange between our universe and out
of its boundary. As a result, we can assume the total
energy-momentum inside the universe is conserved, which implies
$\nabla_{\mu}T^{\mu\nu}=0$. This leads to
\begin{equation}\label{Cont}
\dot{\rho}+3H(\rho+p)=0.
\end{equation}
In addition, due to the volume change of the universe, a work
density term is also appeared as \cite{Hay2}
\begin{equation}\label{Work}
W=-\frac{1}{2} T^{\mu\nu}h_{\mu\nu}=\frac{1}{2}(\rho-p).
\end{equation}
Finally, we write down the first law of thermodynamics on the
apparent horizon as
\begin{equation}\label{FL}
dE = T_h dS_h + WdV.
\end{equation}
Next, we take the total energy inside the apparent horizon as
$E=\rho V$, with $V=\frac{4\pi}{3}R^{3}$ is the volume enveloped
by a 3-dimensional sphere with the area of apparent horizon
$A=4\pi R^{2}$. We now calculate the differential form of the
total energy as,
\begin{equation} \label{dE1}
 dE=4\pi R^{2}\rho dR+\frac{4\pi}{3}R^{3}\dot{\rho} dt.
\end{equation}
Using the continuity equation (\ref{Cont}), we can rewrite the
above equation as
\begin{equation}
\label{dE2}
 dE=4\pi R^{2}\rho dR-4\pi H R^{3}(\rho+p) dt.
\end{equation}
The key point here is to take the entropy associated with the
apparent horizon of FRW universe in the form of Eq. (\ref{Sh3})
with $A=4\pi R^{2}$ is the area of the apparent horizon and $R$ is
the horizon radius. After differentiating, we find
\begin{eqnarray} \label{dSh}
dS_h&=& \frac{dA}{4G}\Big{\{}1+\left(\frac{\gamma
A}{4G}\right)^{\alpha}\Big{\}}^{-1/{\alpha}},
\end{eqnarray}
Substituting relations (\ref{Work}), (\ref{dE2}) and (\ref{dSh})
in the first law of thermodynamics (\ref{FL}) and using definition
(\ref{T}) for the temperature as well as the continuity equation
(\ref{Cont}), after a little algebra, we find the differential
form of the Friedmann equation as
\begin{equation} \label{Fried1}
-2 \left(1+\beta R^{2\alpha}\right)^{-\frac{1}{\alpha}}
\frac{dR}{R^{3}}=
 \frac{8\pi G}{3}d\rho,
\end{equation}
where we have defined
\begin{equation}
\beta\equiv\left(\frac{\gamma \pi}{G}\right)^{\alpha}.
\end{equation}
Next, we integrate Eq. (\ref{Fried1}). The result is
\begin{equation} \label{Frie2}
 \frac{1}{R^{2}} \times \text{{\ }} _{2}F_{1}\Big{\{}
\frac{1}{\alpha},-\frac{1}{\alpha},\frac{\alpha-1}{\alpha}, -\beta
R^{2\alpha}\Big{\}}=\frac{8\pi G}{3}\rho,
\end{equation}
where we have absorbed the integration constant, which can be the
energy density of the cosmological constant, in the total energy
density, namely, $\rho=\rho_m+\rho_{\Lambda}$. Substituting $R$
from Eq.(\ref{radius}), we immediately arrive at
\begin{eqnarray} \label{Fried3}
&&\left(H^2+\frac{k}{a^2}\right)\times \text{{\ }}
_{2}F_{1}\Bigg{\{}
\frac{1}{\alpha},-\frac{1}{\alpha},\frac{\alpha-1}{\alpha}, -\beta
\left(H^2+\frac{k}{a^2}\right)^{-\alpha}\Bigg{\}}=\frac{8\pi
G}{3}\rho.
\end{eqnarray}
In this way, we derive the general form of the modified Friedmann
equation inspired by the MOND theory. The second modified
Friedmann equation can be easily derived by combining Eq.
(\ref{Fried3}) with continuity equation (\ref{Cont}). Using the
fact that $_2F_1(a,b,c,z)$ has a series expansion, we can write
the Friedmann Eq. (\ref{Fried3}) in a compact series form (see
appendix for details),
\begin{eqnarray} \label{Fried3series}
\sum_{n=0}^{\infty}\frac{C_n}{n!}\left(H^2+\frac{k}{a^2}\right)^{1-\alpha
n} =\frac{8\pi G}{3}\rho,
\end{eqnarray}
where
\begin{eqnarray} \label{Cn}
C_n=(-\beta)^n
\frac{\left(\frac{1}{\alpha}\right)_n\left(\frac{-1}{\alpha}\right)_n}{\left(\frac{\alpha-1}{\alpha}\right)_n}.
\end{eqnarray}
Since $\beta<1$, we can expand the hypergeometric function up to
the linear term in $\beta$. This is equivalent to consider the
first and the second term in series (\ref{Fried3series}). We find
\begin{equation} \label{Fried4}
\left(H^2+\frac{k}{a^2}\right)-\frac{\beta}{\alpha
(1-\alpha)}\left(H^2+\frac{k}{a^2}\right)^{1-\alpha} = \frac{8\pi
G}{3} \rho.
\end{equation}
Clearly, the above expression is ill-defined for $\alpha=1$. In
this case, one should start from expression (\ref{ShRenyi}) for
the entropy to derive the modified Friedmann equation. It is easy
to show that for $\alpha=1$, the modified Friedmann equations gets
the following form
\begin{equation} \label{Friedalpha1}
\left(H^2+\frac{k}{a^2}\right)-\frac{\gamma \pi}{G} \ln
\left(H^2+\frac{k}{a^2}\right)=\frac{8\pi G}{3} \rho.
\end{equation}
This is the modified Friedmann equation corresponds to the Reyni
entropy (\ref{ShRenyi}). On the other hand for $\alpha=2$,
expression (\ref{Fried4}) restores
\begin{equation} \label{Friedalpha2}
\left(H^2+\frac{k}{a^2}\right)+\frac{1}{2}\left( \frac{\gamma
\pi}{G}\right)^2 \left(H^2+\frac{k}{a^2}\right)^{-1}=\frac{8\pi
G}{3} \rho.
\end{equation}
This is the modified first Friedmann equation inspired by the
deformed (dual) Kaniadakis entropy (\ref{ShKan}).
\section{Entropic corrections to Newton's law and Friedmann equations \label{Entropic}}
In this section, we are going to apply the idea of entropic
gravity and derive the correction terms to Newton's law of gravity
as well as corrections to Friedmann equations inspired by entropy
expression (\ref{Sh3}). The idea that gravity is not a fundamental
force and can be understood as an entropic force caused by changes
in the information, when a material body moves away from the
holographic screen, was suggested by Verlinde \cite{Ver}.
According to Velinde's proposal when a test particle moves apart
from the holographic screen, the magnitude of the entropic force
on this body has the form
\begin{equation}\label{F}
F\triangle x=T \triangle S,
\end{equation}
where $\triangle x$ is the displacement of the particle from the
holographic screen, while $T$ and $\triangle S$ are the
temperature and the entropy change on the screen, respectively.
Consider two masses, a test mass $m$ and a spherically symmetric
mass distribution $M$ which is surrounded by surface $\mathcal
{S}$. The surface $\mathcal {S}$ is located between  $m$ and $M$
and $m$ is very close to the surface as compared to its reduced
Compton wavelength $\lambda_m=\frac{\hbar}{m}$ ($c=1$). When a
test mass $m$ is a distance $\triangle x = \epsilon \lambda_m$
away from the surface $\mathcal {S}$, the change in the entropy
(\ref{Sh3}) is given by
\begin{equation}
\label{S3}
 \triangle S_h=\frac{\triangle
 A}{4G}\Big{\{}1+\left(\frac{\gamma
A}{4G}\right)^{\alpha}\Big{\}}^{-1/{\alpha}},
\end{equation}
where $A=4\pi R^2$ is the area of the surface $\mathcal {S}$. The
energy inside the surface is identified as $E=M$. On the surface
$\mathcal {S}$, there live a set of ``bytes" of information that
scale proportional to the area of the surface so that that $A=QN$,
where $N$ represents the number of bytes and $Q$ is a fundamental
constant. Note that $N$ is the number of bytes and thus $\triangle
N=1$, hence we have $\triangle A=Q$. According to the
equipartition law of energy, the temperature $T$ in terms of the
total energy on the surface $\mathcal {S}$ reads
\begin{equation}
\label{E}
 T=\frac{2M}{Nk_B}.
 \end{equation}
Substituting Eqs. (\ref{S3}) and (\ref{E}) in Eq. (\ref{F}), we
arrive at
\begin{equation}\label{F3}
F=-\frac{GMm}{R^2}\left(\frac{Q^2}{8\pi k_B \hbar \epsilon
G^2}\right)\Big{\{}1+\left(\frac{\gamma
A}{4G}\right)^{\alpha}\Big{\}}^{-1/{\alpha}}_{A=4\pi R^2},
\end{equation}
where we have assumed the force between $m$ and $M$ is attractive.
In order to arrive at the modified Newton's law of gravity, we
define $Q^2=8\pi k_B \hbar \epsilon G^2$. Taking this into
account, we get
\begin{equation}\label{F4}
F=-\frac{GMm}{R^2}\Big{\{}1+\left(\frac{\gamma
A}{4G}\right)^{\alpha}\Big{\}}^{-1/{\alpha}}_{A=4\pi R^2}.
\end{equation}
Next, we can derive the dynamical equation for the Newtonian
cosmology. We assume surface $\mathcal S$ is the boundary of a
spherical region with volume $V$ and radius $R= a(t)r$ where $r$
is the radial co-moving coordinate. If we combine the second law
of Newton for the test particle $m$ near the surface, with the
gravitational force (\ref{F4}), we reach
\begin{equation}\label{F6}
F=m\ddot{R}=m\ddot{a}r=-\frac{GMm}{R^2}\left[1+\beta
R^{2\alpha}\right]^{-1/{\alpha}}.
\end{equation}
Eq. (\ref{F6}) is nothing, but the modified Newton's law of
gravitation derived by taking the entropy associated with the
holographic screen in the form of (\ref{Sh3}).

The energy density of the matter inside the volume $V=\frac{4}{3}
\pi a^3 r^3$, is  $\rho=M/V$. Thus, Eq. (\ref{F6}) can be
rewritten as
\begin{equation}\label{F7}
\frac{\ddot{a}}{a}=-\frac{4\pi G}{3}\rho\left[1+\beta
R^{2\alpha}\right]^{-1/{\alpha}}.
\end{equation}
This is the modified dynamical equation for Newtonian cosmology.
On the other side, the active gravitational mass is defined as
\cite{Cai4}
\begin{eqnarray}\label{ActM}
\mathcal M &=&2
\int_V{dV\left(T_{\mu\nu}-\frac{1}{2}Tg_{\mu\nu}\right)u^{\mu}u^{\nu}}=(\rho+3p)\frac{4\pi}{3}a^3
r^3.
\end{eqnarray}
Replacing $M$ with $\mathcal M$  ($\rho\rightarrow \rho+3p$)in Eq.
(\ref{F7}), we find
\begin{equation}\label{F8}
\frac{\ddot{a}}{a}=-\frac{4\pi G}{3}(\rho+3p)\left[1+\beta
R^{2\alpha}\right]^{-1/{\alpha}}.
\end{equation}
Multiplying both sides of Eq. (\ref{F8}) with  $2\dot{a}a$ and
using the continuity equation (\ref{Cont}), after integrating we
find
\begin{equation}\label{Frie1}
\dot{a}^2+k=\frac{8\pi G}{3}\int{ d (\rho a^2) \Big{\{}1+\beta
(ra)^{2 \alpha}\Big{\}}^{-1/{\alpha}}},
\end{equation}
where $k$ is an integration constant which can be interpreted as
the curvature constant. To calculate the integral, we first use
the continuity equation (\ref{Cont}), to find
\begin{equation}\label{rho}
\rho=\rho_0 a^{-3(1+w)},
\end{equation}
where $w=p/\rho$ is the equation of state parameter and $\rho_0$
is the present value of the energy density. Substituting relation
(\ref{rho}) in Eq. (\ref{Frie1}), after some calculations, we find
the generalized form of the modified Friedmann equation as
\begin{eqnarray}\label{Frie2}
\left(H^2+\frac{k}{a^2}\right)
\times{_{2}F_{1}\left(\frac{1}{\alpha},-\frac{1+3w}{2
\alpha},\frac{2\alpha-1-3w}{2\alpha}, -\beta
\left(H^2+\frac{k}{a^2}\right)^{-\alpha}\right)}^{-1}=\frac{8\pi
G}{3} \rho.
\end{eqnarray}
Since $\beta\ll1$, we can expand the hypergeometric function to
find the modified Friedmann equation up to the first correction
term. The result is
\begin{equation} \label{Frie3}
\left(H^2+\frac{k}{a^2}\right)-\lambda\left(H^2+\frac{k}{a^2}\right)^{1-\alpha}
= \frac{8\pi G}{3} \rho,
\end{equation}
where
\begin{equation} \label{beta}
\lambda=\frac{\beta(1+3w)}{\alpha (2\alpha-1-3w)}.
\end{equation}
Again, we see that for $\alpha=1$, the second term in Eq.
(\ref{Frie3}) becomes a constant. The result obtained in Eq.
(\ref{Frie3}) from entropic force scenario, is consistent with the
one obtained from the first law of thermodynamics in Eq.
(\ref{Fried4}), which further supports the validity of the entopic
nature of gravity.
\section{Modified Friedmann equations from emergence of cosmic space \label{Emergence}}
In this section, we use the emergence scenario of gravity proposed
by Padmanabhan \cite{PadEm} to find the corrections to the
Friedmann equation based on the modified entropy expression given
in Eq. (\ref{Sh3}). As we have seen in the previous sections, the
correction terms are very small, thus in this section, for
simplicity, we consider the entropy up to the first correction
term given in Eq. (\ref{Sh4}).

According to Padmanabhan, for a pure de Sitter universe with
Hubble constant $H$, the holographic principle can be expressed in
terms of $N_{\rm sur}=N_{\rm bulk}$, where $N_{\rm sur}$, and
$N_{\rm bulk}$, respectively, stand for degrees of freedom on the
boundary and in the bulk. For our real universe, which is
asymptotically de Sitter, as shown by a lot of astronomical
observations, Padmanabhan proposed that in an infinitesimal
interval $dt$ of cosmic time, the increase $dV$ of the cosmic
volume is given by \cite{PadEm}
\begin{equation}
\frac{dV}{dt}\propto
\left(N_{\mathrm{sur}}-N_{\mathrm{bulk}}\right). \label{dV1}
\end{equation}
For a flat universe, Padmanabhan assumed the temperature and
volume as $T=H/2\pi$ and $V=4\pi/3H^3$. The reason for this
assumption comes from the fact that in this case one may consider
our universe as an asymptotically de Sitter space. Padmanabhan
proposed the difference between degrees of freedom on the horizon
and in the bulk, leads to the expansion of our universe.
Mathematically, he assumed \cite{PadEm}
\begin{equation} \label{dV}
\frac{dV}{dt}=G(N_{\mathrm{sur}}-N_{\mathrm{bulk}}).
\end{equation}
Soon after Padmanabhan, his idea was extended to a nonflat
universe by derivation of the Friedmann equations in Einstein,
Gauss-Bonnet and more general Lovelock gravity with any spatial
curvature \cite{Sheyem}. It was argued that in this case one
should replace the Hubble radius ($H^{-1}$) with the apparent
horizon radius $R=1/{\sqrt{H^2+k/a^2}}$, which is a generalization
of Hubble radius for $k\neq0$. The generalization of Eq.
(\ref{dV}), for a nonflat universe was proposed as \cite{Sheyem}
\begin{equation}\label{dV1}
\frac{dV}{dt}=GRH \left(N_{\mathrm{sur}}-N_{\mathrm{bulk}}\right),
\end{equation}
For a flat universe, $RH=1$, and Eq. (\ref{dV1}) restores Eq.
(\ref{dV}). The temperature associated with the apparent horizon
is assumed to be
\begin{equation}\label{T2}
T=\frac{1}{2\pi R}.
 \end{equation}
The reason for taking this expression for temperature instead of
relation (\ref{T}) comes from the fact that here we would like to
consider an equilibrium system, thus within an infinitesimal
internal of time $dt$ we propose $\dot {R}\ll 2HR$, which
physically means that the apparent horizon radius is fixed during
an infinitesimal internal of time $dt$, similar to de-Sitter
Universe. Note that the proposal of Padmanabhan indeed relates the
volume change $dV$ in an infinitesimal interval $dt$ of cosmic
time to the degrees of freedom. Thus it is reasonable to neglect
the dynamical terms in the Hayward surface gravity and approximate
it as $\kappa \simeq {1}/{R}.$ This approximation leads to the
familiar expression for the horizon temperature \cite{CaiEm}.
Besides, since our universe is assumed to be asymptotically de
Sitter, thus one should consider the temperature as (\ref{T2}).
Only with this assumption, one can deduce the correct form of the
Friedmann equations through Padmanabhan's scenario \cite{CaiEm}.
Moreover, in the specific framework of Padmanabhana's emergent
gravity paradigm, relation (\ref{dV}) assumes that the system is
near \textit{thermal equilibrium} at each infinitesimal time step.
In such a setting, taking the horizon radius as effectively
constant during this short interval is physically meaningful and
consistent with the idea of horizon thermodynamics in slowly
varying spacetimes. Note that in section \ref{FIRST}, one can also
consider the temperature associated with the apparent horizon in
the form of (\ref{T2}), however in this case one should apply the
first law as $-dE=TdS$ where $-dE$ is the energy flux crossing the
horizon and the volume term should be absent in the first law of
thermodynamics \cite{Cai2}.

Using the entropy expression (\ref{Sh4}), the number of degrees of
freedom on the surface is given by
\begin{eqnarray} \label{Nsur2}
N_{\mathrm{sur}}&=&4 S_h=\frac{A}{G}\left[1-\eta
\left(\frac{A}{4G}\right)^{\alpha}\right]=\frac{4\pi R^2}{G}+
4\pi\zeta R^{2\alpha+2},
\end{eqnarray}
where $\zeta=-\eta \pi^{\alpha}/G^{\alpha+1}$ and we have taken
$A=4 \pi R^2$ as the area of the boundary. The total energy inside
the apparent horizon is in the form of the Komar energy and is
given by
\begin{equation}
E_{\mathrm{Komar}}=|(\rho +3p)|V,  \label{Komar}
\end{equation}
where  $ V=4 \pi R^3/3$ is the volume of a sphere enveloped by the
apparent horizon. The number of degrees of freedom of the matter
field in the bulk can be obtained using the equipartition law of
energy ($k_B=1$),
\begin{equation}
N_{\mathrm{bulk}}=\frac{2|E_{\mathrm{Komar}}|}{T}.  \label{Nbulk}
\end{equation}
Combining this relation with Eq. (\ref{Komar}) and assuming, in an
expanding universe, $\rho+3p<0$, we find
\begin{equation}
N_{\rm bulk}=-\frac{16 \pi^2}{3}  R^4 (\rho+3p). \label{Nbulk}
\end{equation}
Substituting relations (\ref{Nsur2}) and (\ref{Nbulk}) in
assumption (\ref{dV1}), after simplifying, we arrive at
\begin{eqnarray}
-2\frac{\dot{R}R^{-3}}{H}-2R^{-2}-2\zeta G
R^{2\alpha-2}=\frac{8\pi G }{3}(\rho+3p). \label{Frgb11}
\end{eqnarray}
Next, we multiply both side of Eq. (\ref{Frgb11}) by factor
$-\dot{a}a$, after using the continuity equation (\ref{Cont}), we
reach
\begin{equation}\label{Frgb2}
\frac{d}{dt}\left( a^2 R^{-2}\right)+2\zeta G \dot{a}a
R^{2\alpha-2} =\frac{8 \pi G }{3} \frac{d}{dt}(\rho a^2).
\end{equation}
\\
Taking into account the fact that $R=a(t)r$ in the second term, we
can integrate the above equation. The result is
\begin{equation}\label{Frgb3}
R^{-2}+\frac{\zeta G}{\alpha} R^{2\alpha-2} =\frac{8 \pi G }{3}
\rho.
\end{equation}
Using the fact that $R=1/\sqrt{H^2+k/a^2}$, we finally get
\begin{equation}\label{Frgb4}
\left(H^2+\frac{k}{a^2}\right)+ \chi
\left(H^2+\frac{k}{a^2}\right)^{1-\alpha}=\frac{8\pi G}{3}\rho,
\end{equation}
where $$\chi=\frac{\zeta
G}{\alpha}=-\frac{\beta}{\alpha^2(\alpha+1)}.$$

In this way, we obtain the modified Friedmann equation inspired by
the MOND theory through the method of the emergence gravity. One
can easily check that the result obtained here is consistent with
those obtained in the previous sections from two other approaches,
up to the leading order correction terms. Our studies therefore
further support the viability of the Padmanabhan's perspective of
emergence gravity.
\section{Generalized second law of thermodynamics\label{GSL}}
For a given modified entropy expression associated with the
boundary of the system, one of the main question, which should be
addressed is whether or not the entropy associated with the
horizon can satisfy the generalized second law of thermodynamics.
For an accelerated expanding universe, the generalized second law
of thermodynamics have been investigated in the literatures
\cite{wang2,wang3,SheyGSL}.

Using Eqs.  (\ref{Cont}) and (\ref{Fried1}), we can find
\begin{equation} \label{dotR}
\dot{R}= 4\pi G R^3 H (\rho+p) \left(1+\beta
R^{2\alpha}\right)^{\frac{1}{\alpha}}.
\end{equation}
It is easy to show that
\begin{eqnarray}\label{TSh1}
T_{h} \dot{S_{h}}&=&4\pi H  R^3 (\rho+p)\left(1-\frac{\dot{R}}{2H
R}\right).
\end{eqnarray}
Since our universe is currently experiencing a phase of
accelerated expansion, thus we may have $\rho+p<0$, which implies
the second law of thermodynamics may break down, $\dot{S_{h}}<0$.
Therefore we consider the generalized second law of
thermodynamics. From the Gibbs equation we have \cite{Pavon2}
\begin{equation}\label{Gib2}
T_m dS_{m}=d(\rho V)+pdV=V d\rho+(\rho+p)dV,
\end{equation}
where $T_{m}$ and $S_m$ stand for the temperature and entropy of
the matter fields in the bulk, respectively. We further assume
there is no energy follow between the bulk and the boundary of the
universe. This means that we can take $T_m\approx T_h$
\cite{Pavon2}. Thus from the Gibbs equation (\ref{Gib2}), one
finds
\begin{equation}\label{TSm2}
T_{h} \dot{S}_{m} =4\pi {R^2}\dot {R}(\rho+p)-4\pi {R^3}
H(\rho+p).
\end{equation}
Combining Eqs. (\ref{dotR}), (\ref{TSh1}) and (\ref{TSm2}), one
can arrive
\begin{eqnarray}\label{GSL3}
T_{h}( \dot{S_{h}}+\dot{S_{m}})=8 \pi^2 G H R^{5}(\rho+p)^2
\left(1+\beta R^{2\alpha}\right)^{1/{\alpha}}.
\end{eqnarray}
This confirms that we have always $\dot{S_{h}}+\dot{S_{m}}\geq 0$,
which means that the time evolution of the total entropy,
including the modified entropy associated with the apparent
horizon plus the matter entropy inside the universe is a non
decreasing function of time. This implies that the generalized
second law of thermodynamics holds when the entropy associated
with the apparent horizon is given by Eq. (\ref{Sh3}).
\section{Closing remarks \label{Conc}}
Nowadays, it is a general belief that there is a profound
connection between the laws of gravity and the laws of
thermodynamics. It has been shown that the gravitational field
equations can be derived from thermodynamic arguments in three
levels. In the first level it was shown that the field equations
of gravity can be derived from the first law of thermodynamics. In
a deeper level, it was confirmed that gravity is an entropic
force, which can be understood from statistical mechanics using
two fundamental principles, namely the equipartition law of energy
and the holographic principle. In the deepest level, it was argued
that gravity (geometry) is not a pre-exist quantity and the cosmic
space emerges as the cosmic time progress. This idea leads to
extraction the Friedmann equations describing the evolution of the
FRW universe by counting the degrees of freedom on the boundary
and in the bulk.

In this paper, we have reconsidered thermodynamics-gravity
correspondence to establish a general form of the entropy
associated with the boundary. To this aim, we started from a
general form of the MOND theory and then using an inverse approach
for the entropic force scenario, we reconstructed the general form
of the entropy associated with the boundary. We supposed the
entropy associated with the apparent horizon of FRW universe has
the same expression. This allows us to construct, using three
methods, the modified Friedmann equations by starting from the
modified entropy expression and applying the
thermodynamics-gravity conjecture. We confirmed the consistency of
the obtained results from three approaches, which further supports
the idea of thermodynamics-gravity correspondence.

Based on the modified Friedmann equations derived here, one can
establish a modified cosmological model. Thus, many issues remain
to be addressed. First of all, the cosmological consequences of
the obtained modified Friedmann equations should be studied. In
particular, it is interesting to check whether or not the
cosmological model based on these Friedmann equations can lead to
an accelerated expansion without invoking dark energy. If this is
the case, then we can assert that our model can explain both
challenges of the modern cosmology without invoking unusual
additional component of matter/energy. It is also worthy to
explore cosmological parameters, growth of perturbations,
inflationary models, and the early nucleosynthesis in the context
of the modified Friedmann equations. In addition, one can start
from the general form of the entropy given in this work, and
reproduce the corrections to the Einstein field equations. These
issues are under investigation and the results will be appeared in
the future.
\subsection*{APPENDIX}
The hypergeometric function $_2F_1(a,b,c,z)$ is a mathematic
function which is represented by the hypergeometric series, that
includes many other special functions as specific or limiting
cases. The hypergeometric function $_2F_1(a,b,c,z)$ has a series
expansion as
\begin{eqnarray} \label{Ap1}
&&_2F_1(a,b,c,z)=\sum_{n=0}^{\infty}\frac{(a)_n
(b)_n}{(c)_n}\frac{z^n}{n!}=1+\frac{ab}{c}
\frac{z}{1!}+\frac{a(a+1) b
(b+1)}{c(c+1)}\frac{z^2}{2!}\nonumber\\&&+\frac{a(a+1)(a+2) b
(b+1)(b+2)}{c(c+1)(c+2)}\frac{z^3}{3!}+...,
\end{eqnarray}
where
\begin{equation}\label{Ap2}
(q)_n=\left\{
  \begin{array}{ll}
    1 \quad \quad \quad \quad \quad \quad \quad \quad \quad \quad\!\!{\rm for}\quad $$n=0$$, &  \\
    $$q (q+1)...(q+n-1)$$\quad {\rm for}\quad
    $$n>0$$.
  \end{array}
\right.
\end{equation}
The series terminates if either $a$ or $b$ is a nonpositive
integer, in which case the function reduces to a polynomial:
\begin{eqnarray} \label{Ap3}
_2F_1(-m,b,c,z)=\sum_{n=0}^{\infty}(-1)^n  \frac{m!}{n!(m-n)!}
\frac{(b)_n }{(c)_n}{z^n}.
\end{eqnarray}
The differentiation formula for the hypergeometric function is
\begin{eqnarray} \label{Ap4}
\frac{d^n}{dz^n}\text{{\ }} _2F_1(a,b,c,z)= \frac{(a)_n (b)_n
}{(c)_n}\text{{\ }}  _2F_1(a+n,b+n,c+n,z).
\end{eqnarray}
Many of the common mathematical functions can be expressed in
terms of the hypergeometric function, or as limiting cases of it.
Some typical examples are

\begin{eqnarray} \label{Ap5}
_2F_1(1,1,2,-z)&=&\frac{\ln(1+z)}{z},\nonumber\\
_2F_1\left(\frac{1}{2},\frac{1}{2},\frac{3}{2},z^2\right)&=&\frac{\arcsin(z)}{z},\nonumber\\
_2F_1(a,b,b,z)&=& (1-z)^{-a},  \quad \quad   (b  \ \rm arbitrary).
\end{eqnarray}

\acknowledgments{We thank the anonymous referee for very helpful
and constructive comments which helped us improve our paper. We
are grateful to A. Asvar and A. Shahbazi for useful discussions.
This work is based upon research funded by Iran National Science
Foundation (INSF) under project No. 4022705.}

\end{document}